# A micro channel-cut crystal X-ray monochromator for a self-seeded hard X-ray free-electron laser

**Taito Osaka,[a,*] Ichiro Inoue,[a] Ryota Kinjo,[a] Takashi Hirano,[b] Yuki Morioka,[b] Yasuhisa Sano,[b] Kazuto Yamauchi[b] and Makina Yabashi[a,c]**

[a]RIKEN SPring-8 Center, 1-1-1 Kouto, Sayo-cho, Sayo-gun, Hyogo 679-5148, Japan, [b]Department of Precision Science and Technology, Graduate School of Engineering, Osaka University, 2-1 Yamada-oka, Suita, Osaka 565-0871, Japan, [c]JASRI, 1-1-1 Kouto, Sayo-cho, Sayo-gun, Hyogo 679-5198, Japan. E-mail: osaka@spring8.or.jp

**Synopsis** A channel-cut Si(111) crystal X-ray monochromator with a channel width of 90 μm was developed for achieving reflection self-seeding in hard X-ray free-electron lasers. A high conversion efficiency of a monochromatic seed from a broadband X-ray beam was experimentally evaluated to be ~ $2 \times 10^{-2}$ with a small optical delay of 119 fs at 10 keV.

**Abstract** A channel-cut Si(111) crystal with a channel width of 90 μm was developed for achieving reflection self-seeding in hard X-ray free-electron lasers (XFELs). With the crystal, a monochromatic seed pulse is produced from a broadband XFEL pulse generated in the first-half undulators with an optical delay of 119 fs at 10 keV. The small optical delay allows a temporal overlap between the seed optical pulse and the electron bunch by using a small magnetic chicane for the electron beam at the middle of the undulator section. A peak reflectivity reached 67%, which is a reasonable value as compared with the theoretical one of 81%. By using this monochromator, a monochromatic seed pulse without broadband background in spectrum was obtained at SACLA with a conversion efficiency from a broadband XFEL pulse of ~ $2 \times 10^{-2}$, which is ~ 10 times higher than that of transmission self-seeding using a thin diamond (400) monochromator.

## 1. Introduction

X-ray free-electron lasers (XFELs) operating in the self-amplified spontaneous emission (SASE) scheme (Kondratenko & Saldin, 1980; Bonifacio *et al.*, 1984) have generated hard X-ray beams with unique characteristics, such as ultrahigh peak power, nearly full transverse coherence, and ultrashort pulse duration at the femtosecond level (Huang & Kim, 2007). In the SASE scheme, spontaneous radiation emitted from an electron bunch with a random statistical modulation of the beam density is amplified while propagating through a periodic magnetic field in long undulators. The stochastic nature, however, causes poor longitudinal coherence and a broad spectrum with a relative bandwidth of ~ 0.3% (Saldin, 1999). A monochromator is widely used to monochromatize SASE-XFEL pulses, whereas the photon flux is considerably decreased by a factor of ~ 30.

To generate intense and narrowband XFEL pulses in the hard X-ray regime, a self-seeding scheme has been proposed (Feldhaus *et al.*, 1997). In this scheme, a monochromator is placed at the middle of a magnetic chicane of the electron bunch, which separates the undulators into two sections. A monochromatic seed pulse is produced from a broadband SASE-XFEL pulse generated in the first undulator section through the monochromator. At the same time, the electron bunch is detoured in the magnetic chicane, and then spatiotemporally overlapped with the seed pulse downstream of the chicane. Finally, the seed pulse is amplified in the second undulator section while interacting with the electron bunch. The technical difficulty in self-seeding is to assure a temporal overlap between the electron bunch and the seed pulse. This is because a delay for multi-GeV electron beams is typically limited up to a few hundreds of femtoseconds with a chicane of ~ 5 m length, whereas an optical delay is as large as several picoseconds with a conventional Si(111) double-crystal monochromator (DCM) that has a distance between the crystals of several millimeters. Amann *et al.* demonstrated self-seeding in the hard X-ray regime (Amann *et al.*, 2012) at the Linac Coherent Light Source (LCLS) (Emma *et al.*, 2010) with a thin diamond (400) monochromator operating in the forward Bragg diffraction geometry. Here they used a monochromatic pulse echo as a seeding source that follows a prompt SASE-XFEL pulse with an interval of tens of femtoseconds (Geloni *et al.*, 2010; Geloni *et al.*, 2011; Lindberg & Shvyd'ko, 2012), and successfully generated a narrowband XFEL beam. With the 'transmission' monochromator, however, the prompt broadband SASE-XFEL pulse propagates through an optical axis being common to the seed pulse, which provides a broadband background in spectrum. Furthermore, a conversion efficiency, defined by the ratio of a peak power of a monochromatic seed pulse to that of an input XFEL pulse, of the transmission monochromator has been evaluated to be ~ $10^{-3}$, being much lower than that of a usual 'reflection' monochromator, which could degrade the electron bunch so as to generate an intense seeding source.

In this paper we report development of a micro channel-cut monochromator (Bonse & Hart, 1965) as an essential optical component for achieving the reflection self-seeding with a high efficiency. We designed a channel-cut Si(111) crystal with a channel width of 90 μm, which gives an optical delay as small as ~ 100 fs while removing the prompt SASE beam. Since an X-ray beam is reflected twice at the two opposite blades of the channel-cut crystal made of a *monolithic* single crystal block, the optical axis of the exit beam is highly stabilized as compared with a DCM composed of independent crystals (Diaz *et al.*, 2010). After characterization of quality of the crystal at a SPring-8 beamline, we performed a feasibility test of the reflection self-seeding at SPring-8 Angstrom Compact free-electron LAser (SACLA) (Ishikawa *et al.*, 2012).

## 2. Design of micro channel-cut crystal

To achieve the reflection self-seeding with the micro channel-cut crystal, the following requirements should be satisfied: a small optical delay being less than the maximum delay for the electron bunch, and a large spatial acceptance to receive a major portion of an incident XFEL beam. The delay for the electron bunch $\Delta t_e$ is approximately given by (Hara *et al.*, 2013)

$$\Delta t_e \sim \frac{\theta_e^2}{c}\left(L + \frac{2}{3}L_B\right), \tag{1}$$

where $\theta_e$ is the deflection angle of the electron beam, $c$ the speed of light, $L$ the drift distance between the first and second dipole magnets in the magnetic chicane, and $L_B$ the length of the magnetic field. The deflection angle can be written as

$$\theta_e \sim \frac{cBL_B}{E_e}, \tag{2}$$

where $B$ is the magnetic flux density and $E_e$ the electron beam energy. At SACLA we tune the photon energy of XFEL beams $E_{ph}$ by adjusting the electron beam energy $E_e$ and/or the undulator deflection parameter $K$ according to the following relationship:

$$E_{ph} = \frac{2hc}{\lambda_u}\frac{\gamma^2}{1+K^2/2} \equiv \frac{A\gamma^2}{1+K^2/2}, \tag{3}$$

where $h$ is the Plank's constant, $\lambda_u$ (= 18 mm) the undulator periodic length, and $\gamma$ (= $E_e/E_{e0}$ with the rest energy of the electron $E_{e0}$) the Lorenz factor of the electron beam. The coefficient $A$ (= $2hc/\lambda_u$) is used to simplify the equation. Setting boundary conditions $E_e \leq E_{e\_max}$ ($\gamma \leq \gamma_{max}$) and $K \leq K_{max}$, we have a reference photon energy $E'_{ph} = A\gamma_{max}^2/(1 + K_{max}^2/2)$. For $E_{ph} \leq E'_{ph}$, we usually adjust $E_e$ to tune the photon energy $E_{ph}$ with a fixed $K$ of $K_{max}$, while we change $K$ with fixing $E_e = E_{e\_max}$ for $E_{ph} \geq E'_{ph}$. These conditions give the delay for the electron bunch $\Delta t_e$ as

$$\Delta t_e \sim \begin{cases} C\dfrac{B^2}{E_{ph}} & \left(E_{ph} \leq E'_{ph}\right) \\ DB^2 & \left(E_{ph} \geq E'_{ph}\right) \end{cases} \tag{4}$$

with

$$C = \frac{AcL_B^2\left(L + \frac{2}{3}L_B\right)}{\left(1 + K_{max}^2/2\right)E_{e0}^2}$$

$$D = \frac{cL_B^2\left(L + \frac{2}{3}L_B\right)}{E_{e\_max}^2}. \tag{5}$$

Equation (4) indicates that $\Delta t_e$ is inversely proportional to $E_{ph}$ for $E_{ph} \leq E'_{ph}$, whereas it is independent of $E_{ph}$ for $E_{ph} \geq E'_{ph}$. On the other hand, the optical delay $\Delta t_{ph}$ for a single channel-cut monochromator in the symmetric case is described by

$$\Delta t_{ph} = \frac{hg}{d_{hkl}E_{ph}}, \tag{6}$$

where $g$ is the channel width of the crystal and $d_{hkl}$ the lattice spacing for the ($hkl$) diffraction. In order to introduce the maximum delay $\Delta t_{e\_max}$ to the electron bunch, we apply the maximum magnetic flux density $B_{max}$ to the dipole magnets in the chicane. The upper limit of the width $g_{max}$ is defined that the optical delay with $g = g_{max}$ is equal to $\Delta t_{e\_max}$, which can be written as

$$g_{max} = \frac{d_{hkl}}{h}\Delta t_{e\_max}E_{ph} \sim \begin{cases} C'd_{hkl} & (E_{ph} \leq E'_{ph}) \\ D'd_{hkl}E_{ph} & (E_{ph} \geq E'_{ph}) \end{cases} \tag{7}$$

with $C' = CB_{max}^2/h$ and $D' = DB_{max}^2/h$. Note that $g_{max}$ becomes a constant value $C'd_{hkl}$ for $E_{ph} \leq E'_{ph}$, while it is proportional to $E_{ph}$ for $E_{ph} \geq E'_{ph}$. By using a higher order diffraction with a smaller $d_{hkl}$, the requirement on $g$ becomes more severe, whereas the bandwidth of the seed and the seeded XFEL beam to be achieved could be narrower. The conversion efficiency of the monochromator using a high order diffraction is also reduced as compared with that for a low order diffraction because of the narrower acceptance in spectrum.

Figure 1 shows $\Delta t_{e\_max}$ and the corresponding $g_{max}$ as a function of $E_{ph}$. Here we used the Si(111) diffraction ($d_{111}$ = 3.136 Å) that is the lowest order diffraction in silicon to achieve a high conversion efficiency, and the latest parameters for the magnetic chicane at SACLA ($L$ = 1.56 m, $L_B$ = 0.34 m, and $B_{max}$ = 0.52 T) with boundary conditions $E_{e\_max}$ = 7.8 GeV and $K_{max}$ = 2.15. The reference photon energy $E'_{ph}$ is calculated to be 9.69 keV. We concluded that the requirement on the optical delay is satisfied over a wide photon energy range by using a channel-cut Si(111) crystal with $g < C'd_{111}$ = 204 μm. Note that a smaller electron-beam delay is preferred for suppressing unwanted coherent synchrotron radiation and transporting an electron bunch with higher quality into the second undulator section.

For the micro channel-cut crystal the spatial acceptance is considerably limited. The acceptance in the vertical direction is practically as small as half of $g$, while depending on $E_{ph}$ and quality of the crystal edge. In the horizontal direction the acceptance is determined by the depth of the diffracting blades, which is typically limited up to $10 \times g$ because of the difficulty in fabrication of a narrow trench. A size of the XFEL beams at the middle of the magnetic chicane was evaluated to be ~ 50 μm in full-width at half maximum (FWHM) at 10 keV through a numerical simulation with an FEL simulation code SIMPLEX (Tanaka, 2015). Therefore, a spatial acceptance of > 50 μm in both the directions is needed.

We designed a prototype micro channel-cut crystal, as shown in Fig. 2(a), which satisfies all of the above requirements near 10 keV. The channel width $g$ was set to 90 μm, which gives an optical delay of 119 fs at 10 keV, as shown in Fig. 1. The overlap length of the two diffracting blades was designed to be 0.2 mm so as to obtain a spatial acceptance in the vertical direction of, at least, 50 μm while avoiding the use of an area near the crystal edge. The depth of the channel is ~ 0.6 mm being sufficiently larger than the horizontal beam size. The dimensions of the crystal are $10 \times 10 \times 10$ mm$^3$ to facilitate handling of the crystal while the right upper part is removed to place a beam stopper just downstream of the first blade, which is required to block the unwanted transmission beam propagating through the crystal. We note that a single channel-cut monochromator imposes a transverse shift of the beam axis $\Delta h$ described by

$$\Delta h = 2g\cos\theta_B,  \quad (8)$$

where $\theta_B$ is the Bragg angle. The shift is less than 180 μm with the micro channel-cut crystal of $g = 90$ μm, which is acceptable for all the optical components at SACLA. Although the shift can be compensated by using a double channel-cut monochromator operating in the (+, −, −, +) geometry, the optical delay becomes twice as large as that in the single channel-cut monochromator.

The micro channel-cut crystal was made of a floating-zone Si single crystal block. An ultra-narrow trench with 60–70 μm width and ~ 0.6 mm depth was formed with a dicing blade. Then the surface layer of 10–15 μm depth was removed through solution etching. Finally, the working area was moderately polished with fine slurry. Since the channel width of the crystal is one of the most important parameter and difficult to be measured with an optical microscope, we accurately characterized the channel width with an X-ray beam through Eq. (8), as described in the section 3.2.

The crystal was mounted on a dedicated holder (Fig. 2(b)) with a beam stopper made of tungsten of 250 μm thickness. The position of the crystal was finely adjusted by positioning screws so as to fit the incident surface to the bottom surface of the beam stopper, as shown in Fig. 2(c).

3. Characterization of micro channel-cut crystal

3-1. Experimental setup

We characterized the micro channel-cut crystal through X-ray reflection topography and rocking curve measurements at the 1-km-long beamline BL29XU of SPring-8 (Tamasaku et al., 2001). A 10 keV X-ray beam emitted from an undulator was monochromatized by a SPring-8 standard Si(111) DCM (Yabashi et al., 1999). The size of the X-ray beam was reduced to 500 × 50 μm$^2$ in the horizontal (H) and vertical (V) directions, respectively, with a 4-jaw slit located 987 m downstream of the source. The rocking curve was measured with two ion chambers, while the reflection topographs were taken with a high-resolution X-ray camera with a pixel size of 325 nm (Kameshima et al., 2018) placed at ~ 1 m downstream of the crystal.

3-2. Results

Figure 3(a) shows a reflection profile under the Bragg condition. Both the incident and reflected beams were observed in a field of view of the camera (~ 665 × 665 μm$^2$), simultaneously. A slightly curved reflection profile was observed, which was due to the variation of the channel width along the horizontal direction. The variation of Δ$h$, however, was less than a few micrometers except near the crystal edge, which could provide a negligible variation of the optical delay. From Eqs. (6) and (8) with the shift Δ$h$ = 176 μm at 300 μm far from the edge, the channel width $g$ and the corresponding optical delay Δ$t_{ph}$ at 10 keV were evaluated to be 90.0 μm and 119 fs, respectively, which well agrees with the design value. The spatial acceptance, defined by the maximum size of the reflected beam, was evaluated to be ~ 600 (H) × 100 (V) μm$^2$ by using a larger X-ray beam.

Figure 3(b) shows a rocking curve to be measured. The peak reflectivity reached 67% while the ideal reflectivity was calculated to be 81%. The reduction in reflectivity was associated with residual surface roughness and damage in the illumination area of the crystal. The imperfection of the crystal also produced speckles on the reflected beam, as shown in Fig. 3(a). We assumed that the speckles originated from the modulation of phase due to the surface undulation. The phase shift $\Delta\varphi$ caused by the variation in surface height $h_S$ is then given by

$$\Delta\phi = \frac{4\pi\delta h_S}{\lambda \sin\theta_B}, \qquad (9)$$

where $\delta$ is the deviation from unity of the real part of the refractive index $n$ (= $1 - \delta + i\beta$). Figure 4(a) shows a distribution of phase shift calculated from the surface morphology on the crystal measured with an optical microscope. A large phase shift of $\sim \pi/2$ with high spatial frequencies of $\sim 65$ mm$^{-1}$ could be occurred within the footprint of the X-ray beam. We consider that this assumption is reasonable because the spatial frequency of the distribution of phase modulation is consistent with that of the speckles. To evaluate the influence of the distorted wavefront on the self-seeding process, we performed an FEL simulation with $E_e = 7.8$ GeV and $K = 2.1015$ ($E_{ph} = 10$ keV) under simple conditions as follows: (1) a monochromatic seed pulse is produced from a broadband SASE pulse with a simple bandpass filter located at the exit of the magnetic chicane (without optical delays), (2) the phase modulation is introduced to the seed pulse after the bandpass filter, (3) the seed pulse and electron bunch start interaction with each other after propagating through 6 m free space, and (4) the electron bunch is perfectly overlapped with the seed pulse in both time and space when the phase modulation is not introduced. Other parameters were not optimized to focus on the influence of the phase modulation. Figure 4(b) shows a comparison between simulated gain curves with and without the phase modulation. This simulation implies that the distorted wavefront of the seed pulse does not provide serious degradation in the amplification process, possibly because the phase modulation with high spatial frequencies is smeared while propagating through the second undulator section, and the TEM$_{00}$ spatial mode in the seed pulse is preferentially amplified. Note that this simulation did not take into account the reduction in peak reflectivity, which could decrease the spectral brightness of the seed and the resultant seeded XFEL beam while providing a similar trend of the gain curve to this simulation.

## 4. Feasibility test at SACLA

We tested the micro channel-cut monochromator at BL3 of SACLA (Tono *et al.*, 2017) in which 21 undulator modules (U1–U21) are implemented. The magnetic chicane with a length of 5.1 m is placed between U8 and U9. A SASE XFEL beam needed for producing a monochromatic seed was generated with five undulator modules (U4–U8) in the first undulator section. The averaged peak power of the SASE XFEL beam was estimated to be 4.8 GW (48 μJ within a 10 fs pulse width) at 10 keV with a bandwidth of ~ 30 eV in FWHM. The micro channel-cut monochromator was placed at the middle of the magnetic chicane. The size of the SASE beam at the monochromator was evaluated to be ~ 65 (H) × 45 (V) μm$^2$ in FWHM through knife-edge scans with the beam stopper, which was consistent with the results of the FEL simulation. Figure 5 displays the averaged spectrum of the seed without amplification in the second undulator section measured by scanning a speckle-free Si(220) channel-cut crystal (Hirano *et al.*, 2016) placed in the experimental hutch 2, 167 m downstream of the monochromator. The spectrum shows a nearly ideal bandwidth of ~ 1 eV in FWHM without broadband background, indicating that the crystal successfully produced a seed beam while removing the transmitted SASE beam. With the peak power of the seed of ~ 88 MW (0.88 μJ, 10 fs), the conversion efficiency was experimentally estimated to be ~ 2 × 10$^{-2}$, which is ~ 10 times higher than that of the transmission monochromator to be reported previously.

## 5. Conclusion and future perspectives

In conclusion, we have developed a micro channel-cut Si(111) crystal monochromator for self-seeding in hard X-ray FELs. The channel width was evaluated to be 90.0 μm, which gives an optical delay of 119 fs to a 10 keV seed pulse. A clean seed beam without broadband background in spectrum was obtained through the micro channel-cut monochromator at BL3 of SACLA. An unprecedented high conversion efficiency from a broadband SASE beam to a seed of ~ 2 × 10$^{-2}$ was experimentally confirmed, allowing one to use an original SASE beam at moderate intensities, and therefore, a high-quality electron bunch can be used for amplification of the seed.

Although the clean and intense seed pulse to be available with the micro channel-cut monochromator and the resultant high-quality electron bunch should increase the spectral brightness of seeded XFELs, further enhancement of spectral brightness will be achieved if the surface quality of the crystal is improved. Chemical etching with atmospheric-pressure plasma employed for fabrication of speckle-free crystal optics (Osaka *et al.*, 2013; Hirano *et al.*, 2016), called plasma chemical vaporization machining (PCVM) (Mori *et al.*, 2000), is a promising approach for this purpose. Furthermore, a seeded XFEL beam generated with the micro channel-cut monochromator using the Si(111) diffraction will not be fully coherent in time because the bandwidth of the seed (~1 eV) is wider than the typical spike width of the SASE beam (~ 0.4 eV) (Inubushi *et al.*, 2012; Inubushi *et al.*,

2017; Osaka *et al.*, 2017). Generation of fully coherent XFEL pulses could be realized by using a high order diffraction of which bandwidth is narrower than the spike width. Achieving the fabrication of the speckle-free micro channel-cut crystal through the PCVM technique allows ones to use a high order diffraction with nearly ideal conversion efficiency and spectral brightness of the seed, which should pave the way to realization of fully coherent XFELs with high intensities.

**Acknowledgements**

The authors would like to thank the staff of SPring-8 and SACLA facilities for their continuous support. We would also like to thank Dr. Y. Kohmura for his support in the characterization of the micro channel-cut crystal at BL29XU of SPring-8, and Drs. T. Hara, T. Inagaki, H. Ohashi and Y. Inubushi for their help for installing the monochromator to BL3 of SACLA. We are grateful for the valuable advices from Drs. H. Tanaka, S. Goto, K. Tamasaku, T. Tanaka, K. Togawa, K. Tono and T. Ishikawa. The characterization of the crystal was carried out with the approval of RIKEN (Proposal Nos. 20170003 and 20180061).

**Funding information**

The following funding is acknowledged: Special Postdoctral Researcher Program of RIKEN; JSPS KAKENHI Grant No. 18K18307.

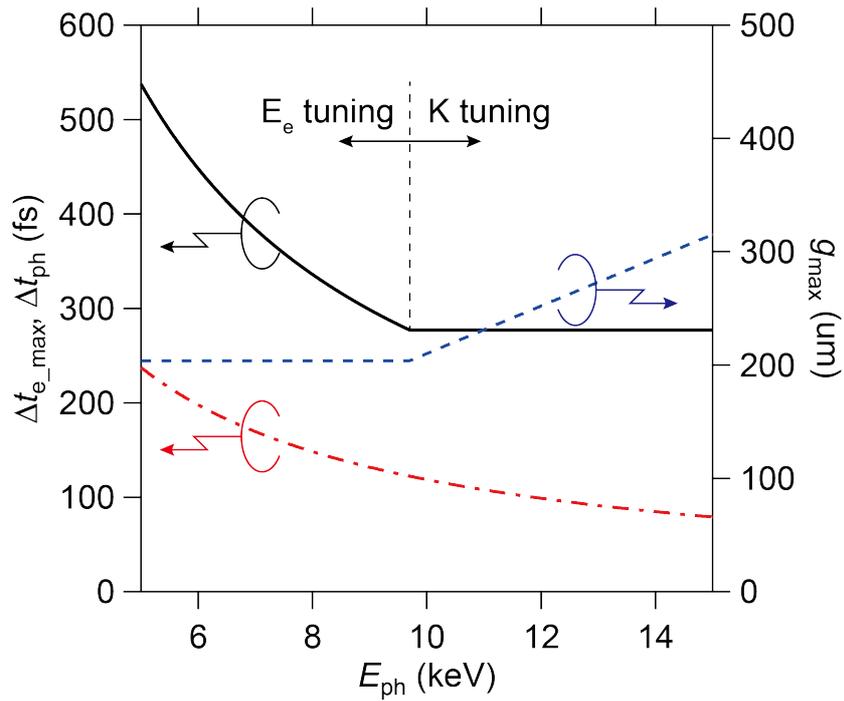

**Figure 1** The maximum delay for the electron beam $\Delta t_{e\_max}$ (solid line) introduced with the magnetic chicane at BL3 of SACLA and the corresponding channel width $g_{max}$ (dashed line) as a function of photon energy $E_{ph}$. Chain line shows the optical delay imposed by the micro channel-cut monochromator with $g = 90$ μm.

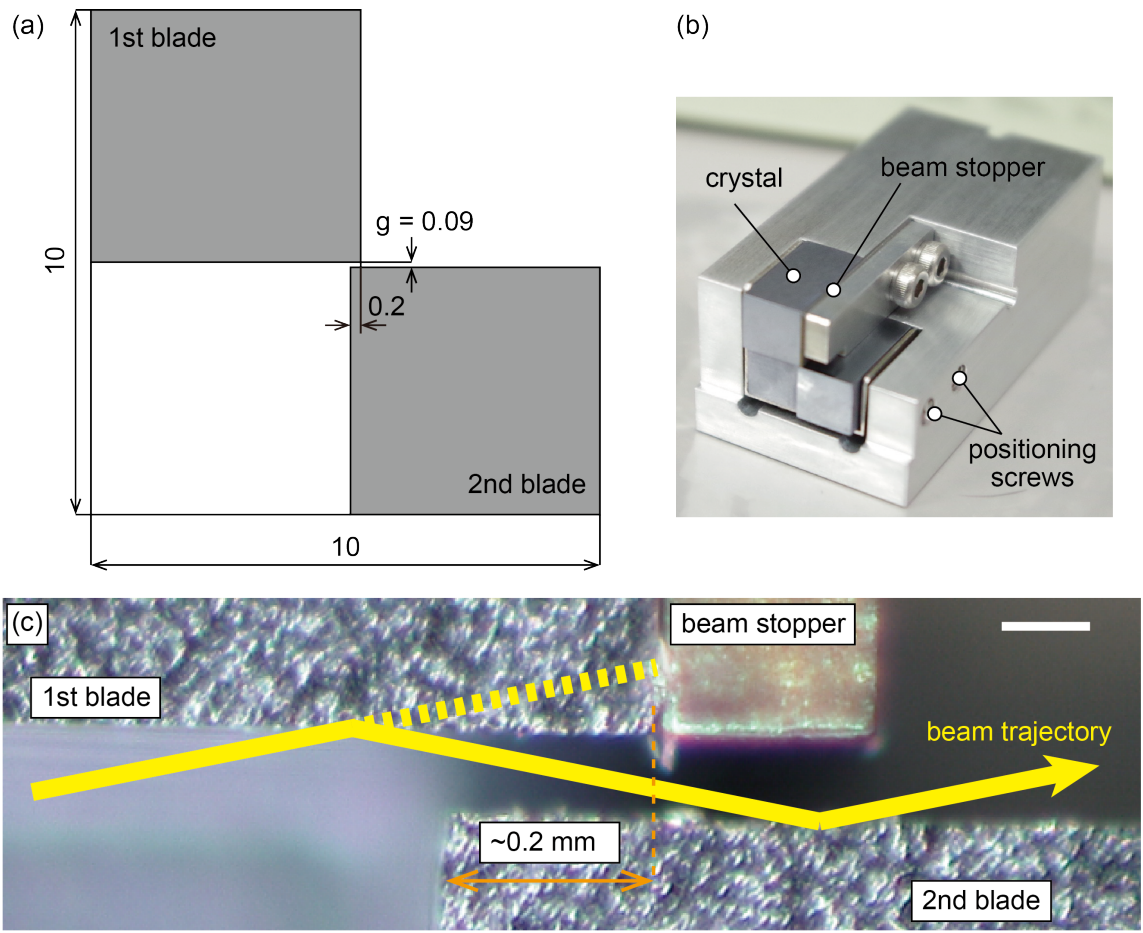

**Figure 2** (a) Drawing of a micro channel-cut crystal viewed from perpendicular to the optical axis. Unit is mm. (b) Picture of the crystal mounted on a dedicated holder. X-ray beams propagate from the left to right side. The position of the crystal is finely adjusted by pushing a stainless plate on the crystal from the left, right and bottom sides with screws so as to fit the surface of the 1st blade to the bottom surface of the beam stopper, as shown in (c). (c) Optical microscope image of the working area. Scale bar shows 100 μm.

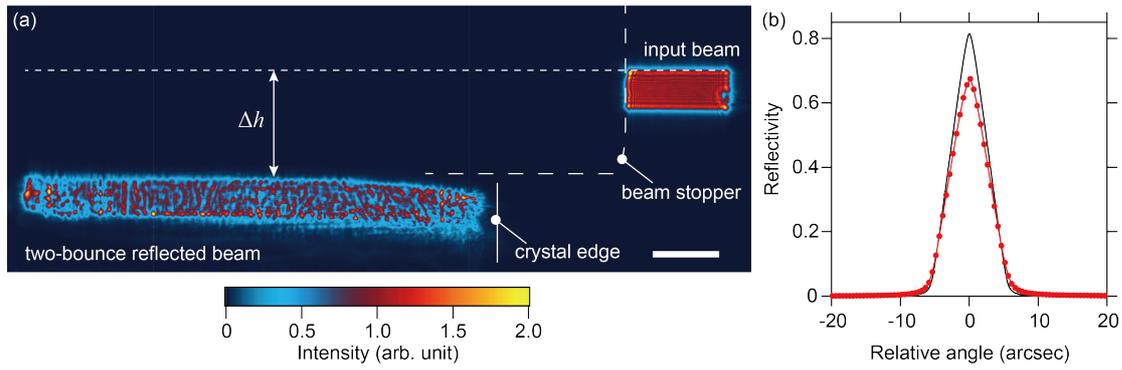

**Figure 3** (a) Measured two-bounce reflection profile from a micro channel-cut crystal under the exact Bragg condition. This image is created by combining some images taken at several crystal positions. The upper-right beam is the incident beam propagating out of the crystal and beam stopper. The location of the beam stopper and the crystal edge is displayed as long-dashed line and solid line, respectively. Scale bar shows 100 μm. (b) Rocking curves measured (plots) and calculated (solid line).

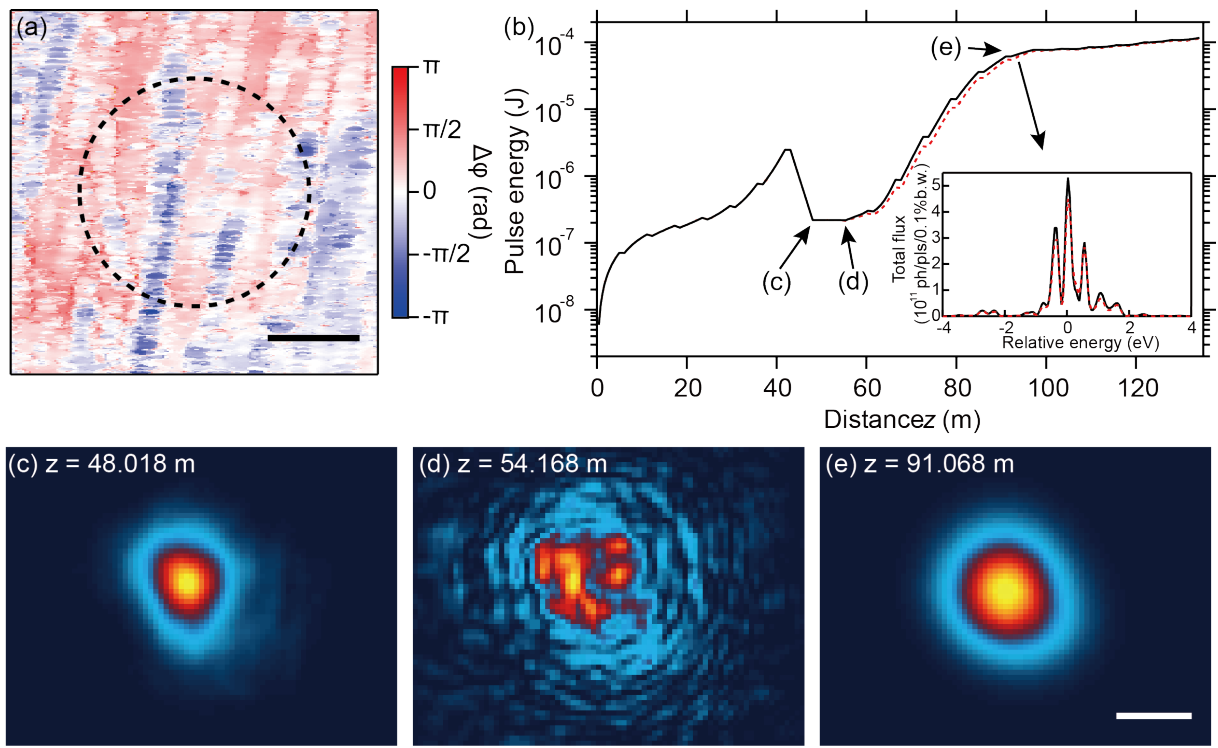

**Figure 4** (a) Distribution of phase shift converted from the surface morphology on the crystal measured with an optical microscope. Dashed circle shows an expected footprint of XFEL beam with a diameter of 50 μm. Scale bar represents 20 μm. (b) Simulated gain curves in the reflection self-seeding using a seed without (solid line) and with the phase modulation (dashed line). The electron beam energy is 7.8 GeV with $K$ = 2.1015 ($E_{ph}$ = 10 keV). Inset of (b) shows an example of the spectrum at a distance $z$ = 91.1 m. Spatial profiles of the XFEL beam at (c) $z$ = 48.0 m (just downstream of the bandpass filter), (d) $z$ = 54.2 m where the seed pulse starts interaction with the electron bunch, and (e) $z$ = 91.1 m. Scale bar represents 50 μm.

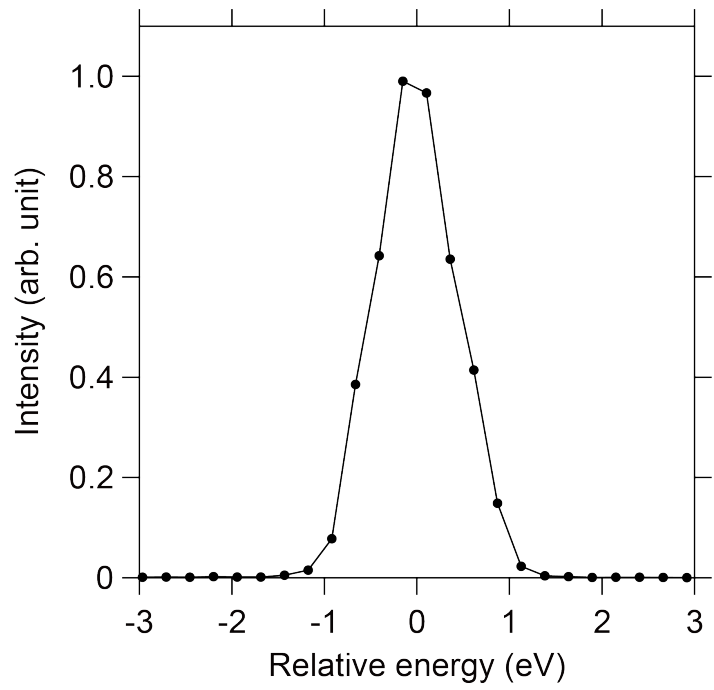

**Figure 5** Averaged spectrum of the seed measured with a Si(220) channel-cut crystal. The number of accumulation at each point is 100 shots. The central photon energy is 10 keV.